%% file: main.tex
\documentclass{article}

\usepackage{amsmath, amsthm, amssymb, amsfonts}
\usepackage{thmtools}
\usepackage{graphicx}
\usepackage{setspace}
\usepackage{geometry}
\usepackage{float}

\usepackage[utf8]{inputenc}
\usepackage[english]{babel}
\usepackage{framed}
\usepackage{svg}

\usepackage{xcolor}

\usepackage{colortbl}
\usepackage{longtable}
\usepackage{array}
\usepackage{color}
\usepackage{tabularray}
\usepackage{breakcites}
\usepackage{hyperref}

\newcommand{\HRule}[1]{\rule{\linewidth}{#1}}

\begin{document}


\title{ \normalsize \textsc{}
		\\ [2.0cm]
		\HRule{1.5pt} \\
		\LARGE \textbf{A Systematic Review on the Potential of AI and ChatGPT for Parental Support and Child Well-Being}
		\HRule{1.5pt} \\ [0.6cm] \normalsize{Technical Report} \vspace*{10\baselineskip}}
		
\date{}
\author{\textbf{Author} \\ 
		Mohsena Ashraf \\
		PhD student in Computer Science \\
            University of Colorado Boulder\\
            May 2024
		}

\maketitle
\newpage

\begin{abstract}
This research article explores the potential of Artificial Intelligence (AI) and Chat Generative Pre-trained Transformers (ChatGPT) in revolutionizing the landscape of parental support and child care. With the advancement of generative AI, conversational agent technology, and Large Language Models, these tools can be a great support for parents to guide and assist their children. Recognizing the challenges faced by parents in navigating the complexities of child-rearing, this study seeks to explore the applications of AI, particularly leveraging the capabilities of ChatGPT, to provide valuable assistance and guidance. For this purpose, we perform a comprehensive examination of the existing 27 literature studies to investigate the potential of AI-driven platforms for offering advice tailored to the individual needs of the parents. We break down these research works into three domains based on their research criteria, and categorize each domain into two subgroups: pre-chatGPT era, and post-chatGPT era. At last, we investigate the significant dimensions of these literatures and present some valuable findings and future research directions.
\end{abstract}
\newpage

\tableofcontents
\newpage

\input{Intro.tex}

\input{Method.tex}
\input{Theme1}
\input{Theme2}
\input{Theme3}
\input{Discussion}
\input{Conclusion}
\newpage
\bibliographystyle{apalike}
\bibliography{References.bib}



\end{document}

%% file: Intro.tex
\section{Introduction}

With the recent advancement of Artificial Intelligence (AI), Generative AI, and Large Language Models (LLMs) like chatGPT, a new era of possibilities in various domains has emerged \cite{intro_1_DeAngelis2023}. One such promising avenue is the realm of parenting, where the integration of these technologies offers innovative solutions and support mechanisms for parents in their crucial role of nurturing and guiding their children. This study dives deeper into the potential applications and benefits of leveraging AI, and ChatGPT, to assist parents in navigating the complexities of modern parenting. Through a comprehensive examination of existing literature and emerging trends, we seek to provide insights into the ways AI and ChatGPT can be harnessed as valuable tools for parents in the evolving landscape of child-rearing.

For this purpose, we have selected 27 papers concentrating on the beneficial applications of using AI and chatGPT for parenting and skill development of children. We systematically categorized these papers into the following three principal domains, each addressing distinct aspects of the interaction between AI and parenting:

\begin{figure}[htb]

    \center
     \includesvg[width=0.9\textwidth]{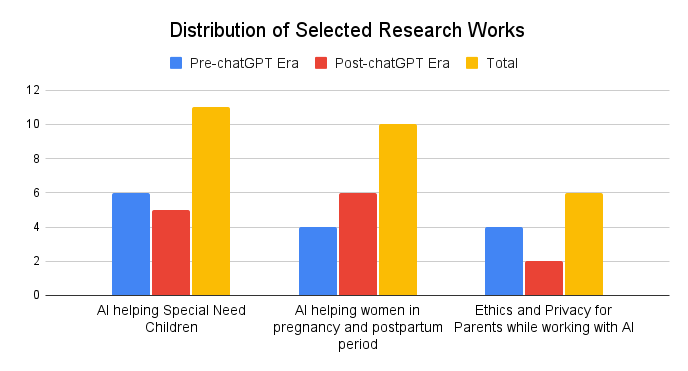}
    \caption{Distribution of Selected Research Works.}
    \label{fig:dist_works}
\end{figure}

\begin{itemize}
    \item \textbf{AI helping Special Need Children (Domain 1):}
    This domain addresses the utilization of AI to assist children with special needs, such as- including those with Autism Spectrum Disorder (ASD), Neurodevelopmental Disorder (NDD), or various physical disabilities. This domain encompasses a compilation of 11 research articles.
    \item \textbf{AI helping women in pregnancy and postpartum period (Domain 2):}
    This domain scrutinizes previous research studies exploring the role of AI in addressing queries and providing suggestions during pregnancy, as well as its potential in mitigating postpartum depression. A total of 10 research papers contribute to this domain.
    \item \textbf{Ethics and Privacy for Parents while working with AI (Domain 3):}
    While engaging with AI, parents and children both, especially parents need to be concerned about the aspect of ethics and privacy of using these technologies. This domain comprises 6 research papers devoted to investigating the ethical implications and privacy concerns inherent in the use of AI.
\end{itemize}

Furthermore, each of these three domains is delineated into two subgroups: pre-chatGPT era and post-chatGPT era. In the pre-chatGPT era, the focus is on discerning the advantages offered by various applications, deep learning-based models, and conversational agents within specific domains. Conversely, in the post-chatGPT era, the emphasis shifts towards exploring the efficacy of generative AI, ChatGPT or other large language models in addressing the needs of individuals in these three domains. Figure \ref{fig:dist_works} visually represents the distribution of the selected papers, providing a comprehensive overview of their alignment with the study's objectives.

%% file: Method.tex
\section{Method of Literature Selection and Analysis}
\label{Method}

\begin{figure}[htb]

    \center
     \includesvg[width=0.9\textwidth]{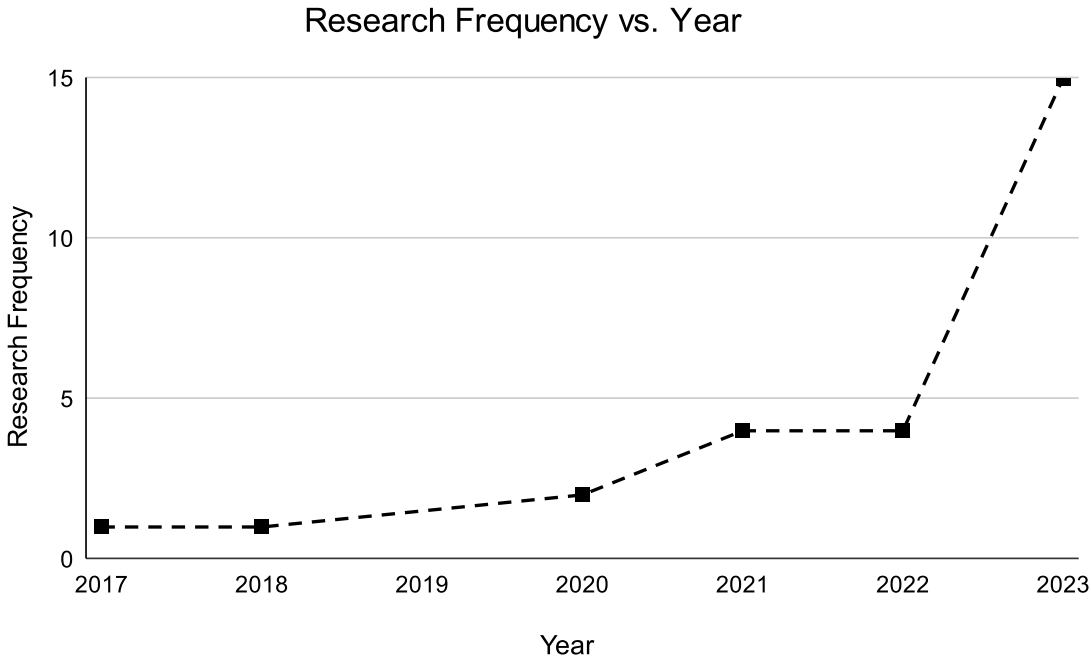}
    \caption{Year-wise Research frequency.}
    \label{fig:work_vs_year}
\end{figure}

Initially, we selected 41 papers in total for our research purpose (13 papers in domain 1, 15 papers in domain 2, and 13 papers in domain 3). The selection of the papers was based on recent publications, the contributions of the authors, and their focus on AI, LLMs, and parenting. But we had to exclude some of the papers due to their nature of being a review or survey paper, as we are already working on presenting a literature review. After the exclusion process, we are finally left with 27 papers ranging from 2017 to 2023 to perform a thorough analysis on what has been done before, and what more can be done in future. The yearwise research frequency can be found in Figure \ref{fig:work_vs_year}.

For each domain, we subdivided the papers into two categories: pre-chatGPT and post-chatGPT era. For domain 1 which focuses on aiding special needs children, in the pre-chatGPT era, we have selected 6 papers, and for the post-chatGPT era, we have selected 5 papers. In domain 2 focusing on assisting women in pregnancy and postpartum period, pre-chatGPT era consists of 4 and post-chatGPT era consists of 6 papers. And for the last domain concentrating on the ethics and privacy, pre-chatGPT era comprises of 4 papers, and post-chatGPT comprises of only 2 papers. Here, we only select 2 papers, because we have found that there is a significant research gap in exploring the ethical aspects of using chatGPT or any other LLMs in terms of parenting. Rather, researchers focused on how to incorporate chatGPT ethically in academia and education.

After the selection of the research works, we pointed out several features of these papers, and upon performing a deep dive investigation, we filled out their values. For all the three domains, we try to dig out the most common 33 features. For domain 1, we add 2 more features, while for domain 2 and 3, we augment 1 features each respectively. The name of the features that has been used in each domain can be found in Table \ref{tab:table-0}. 

\input{table-0}

We incorporated two ways to represent our findings from the literature review. First, we select five significant features for all the papers in each domain, and analyze the studies based on them, which can be found in section $3,4,$ and $5$. After that, for each domain, we select 4 papers (2 from pre-chatGPT and 2 from post-chatGPT era) from domain 1 and 2, and present a brief overview of them based on all the dimensions listed on Table \ref{tab:table-0}. For domain 3, we only select 2 papers (1 from pre-chatGPT and 1 from post-chatGPT era) and present their brief overview in a similar way. We selected only 2 papers in this domain as it contained less papers due to the shortage of targeted research.


After the thorough analysis, we have found that there are still opportunities to explore in these domains. We present our findings after the analysis of each feature in every domain. A brief summary of these findings can be found also in the Section \ref{discussion}.

%% file: table-0.tex
\begin{longtblr}[
  caption = {Selection of features for each domain.},  label={tab:table-0}
]{
  width = \linewidth,
  colspec = {Q[44]Q[229]Q[665]},
  cells = {c},
  cell{2}{2} = {r=33}{},
  cell{35}{2} = {r=2}{},
  vlines,
  hline{1-2,35,37-39} = {-}{},
  hline{3-34,36} = {1,3}{},
}
\textbf{SL} & \textbf{Domain}    & \textbf{Feature Name}                                    \\
1           & Domain 1, 2, and 3 & Year                                                     \\
2           &                    & Main task                                                \\
3           &                    & Diversity of the participants~                           \\
4           &                    & Evaluation method (interview/no evaluation/others)       \\
5           &                    & Method of interview                                      \\
6           &                    & Model used (ChatGPT 3.5/4/others)                        \\
7           &                    & Number of responses generated by ChatGPT                 \\
8           &                    & Department of the paper coming from                      \\
9           &                    & Concerns/challenges of AI mentioned~                     \\
10          &                    & Environment (Home/school/hospital)                       \\
11          &                    & Country of the study                                     \\
12          &                    & Empirical/theoretical                                    \\
13          &                    & Duration of the study                                    \\
14          &                    & Number of sessions                                       \\
15          &                    & IRB approval (yes/no)                                    \\
16          &                    & Informed consent of the participants                     \\
17          &                    & Number of participants                                   \\
18          &                    & Method of participant recruitment                        \\
19          &                    & Target age                                               \\
20          &                    & Actors of the study                                      \\
21          &                    & Who are helpers? (parents/health professionals/teachers) \\
22          &                    & Award for participation                                  \\
23          &                    & Type of data used                                        \\
24          &                    & Classifier/machine learning model used                   \\
25          &                    & Conversational agent (yes/no)                            \\
26          &                    & If conversational agent, which one?                      \\
27          &                    & Voice customization of the CA?                           \\
28          &                    & Citations                                                \\
29          &                    & Metrics used in the study                                \\
30          &                    & Number of the requests to the voice assistant            \\
31          &                    & History logs used? (yes/no)                              \\
32          &                    & Number of misinterpreted commands                        \\
33          &                    & Publishing journal/conference                            \\
34          & Domain 1           & Type of skill (social skill/others)                      \\
35          &                    & Type~ of disability condition                            \\
36          & Domain 2           & Development of an app                                    \\
37          & Domain 3           & Providing suggestions based on findings                  
\end{longtblr}

%% file: Theme1.tex
\section{Analysis of Domain 1 - AI helping Special Need Children:}
\label{theme:1}
In this section, we present our findings for Domain 1. Table \ref{tab:table-1} represents the significant feature values in a structured manner. We have selected the features - \textit{Diversity of the participants, Evaluation method, Model of conversational agent or chatGPT, Participant recruitment strategy, and type of disability and skill development} as our significant features in this particular domain.

\input{table-1}

Now, we elaborately discuss these significant feature dimensions for all of the 11 papers in this particular domain with the analysis implications between the pre and post-chatGPT era:

\begin{enumerate}
    \item \textbf{Diversity of the Participants:}
    
    In the pre-chatGPT era, none of the papers mentioned the diversity of the participants in terms of demography or talked about underrepresented communities. However, the authors in \cite{articleGarg} directed their attention toward collaborating with individuals presenting diverse forms of disabilities, while the authors of \cite{5_10.1145/3271484} delineated various learning difficulties and conditions within the Autism Spectrum.

    Whereas in the post-chatGPT era, only 1 paper \cite{7_10.1145/3585088.3593867} mentioned the inclusiveness of the participants, engaging individuals from both the United States and South Korea. Researchers of the paper \cite{11_Bushuven2023} recruited no human participants, rather they themselves tested their designed vignettes. The remaining papers did not incorporate participants from diverse backgrounds.

    In both groups, working with diverse or underserved participants was not common. However, in the post-chatGPT era, we can see that one paper explicitly addressed inclusiveness and worked with international participants.

    \item \textbf{Evaluation Method used in the Research study:}

    Researchers in most of the studies \cite{1_10.1145/3531073.3531157}, \cite{2_10.1145/3411763.3451666}, \cite{articleGarg}, \cite{4_10.1145/3411764.3445290}, \cite{6_10.1145/3411764.3445116} have chosen interviews as their evaluation method. Among these, \cite{1_10.1145/3531073.3531157} and \cite{2_10.1145/3411763.3451666} opted for a semi-structured interview format, while \cite{articleGarg} employed focused interviews. In addition to interviews, \cite{1_10.1145/3531073.3531157}, \cite{2_10.1145/3411763.3451666}, and \cite{5_10.1145/3271484} incorporated online questionnaire forms, with \cite{articleGarg} also utilizing content analysis for evaluation. Authors in \cite{4_10.1145/3411764.3445290} integrated observational notes and email feedback from participating teachers in their study. Furthermore, \cite{5_10.1145/3271484} utilized the British Picture Vocabulary Scales (BPVS) to assess children's specific difficulties and implemented the Social Communication, Emotional Regulation, and Transactional Support (SCERTS) model. Regarding interview methods, \cite{1_10.1145/3531073.3531157} and \cite{2_10.1145/3411763.3451666} conducted online interviews using Zoom and Google Meet, respectively, while \cite{6_10.1145/3411764.3445116} conducted interviews through both phone and face-to-face sessions. Other studies did not specify the interview conduct methodology.

    In the post-ChatGPT era, only \cite{7_10.1145/3585088.3593867} utilized audio interviews conducted via Zoom as their chosen evaluation method. Conversely, \cite{8_Imran2023} and \cite{9_Yang2023} did not employ specific evaluation methodologies in their respective studies. Notably, researchers in the investigation conducted by \cite{10_Moise2023} undertook a comparative analysis of ChatGPT-generated responses with recommendations outlined in the most recent American Academy of Otolaryngology–Head and Neck Surgery Foundation (AAO-HNSF) "Clinical Practice Guideline: Tympanostomy Tubes in Children (Update)" through the evaluation of two otolaryngologists. Additionally, in \cite{11_Bushuven2023}, the authors developed and validated their vignettes through the collaborative assessment of five emergency physicians.

    In the domains investigated during the pre-chatGPT era, the prevalent evaluation methods in research studies were interviews and participant feedback. These interviews were often conducted using online platforms such as Zoom and Google Meet. Conversely, in the post-chatGPT era, the assessment of chatGPT-generated responses saw a shift. Physicians reviewed the responses, and comparisons were made with established medical guidelines. Additionally, some studies solely deliberated on the potential of chatGPT without direct implementation.

    \item \textbf{Model of Conversational Chatbot or ChatGPT used in the study:}

    In the pre-ChatGPT era, most of the empirical studies employed conversational agents to assist children with special needs. For example, \cite{1_10.1145/3531073.3531157} utilized the 3rd generation Amazon Alexa Echo Dot, \cite{2_10.1145/3411763.3451666} employed Google Home with Google Assistant, and \cite{6_10.1145/3411764.3445116} implemented Nugu Candle, a popular AI assistant in Korea. In contrast, \cite{5_10.1145/3271484} employed a virtual character named ANDY, \cite{4_10.1145/3411764.3445290} utilized an augmented reality (AR) device named PeopleLens, specifically using HoloLens, and \cite{articleGarg} did not incorporate any AI assistant, as it was a theoretical study.

    The studies which employed chatGPT in their work mostly used GPT-3 \cite{7_10.1145/3585088.3593867} or GPT-3.5 \cite{10_Moise2023}. Authors of \cite{7_10.1145/3585088.3593867} also utilized Dall.E 2, Midjourney, and Stable Diffusion in their research. Only \cite{11_Bushuven2023} used the latest version of chatGPT powered by GPT-4. On the other hand, \cite{8_Imran2023}, and \cite{9_Yang2023} only discussed the benefits of incorporating chatGPT to promote child mental health care and supporting their cognitive development.

    It is evident that in the pre-chatGPT era, the prevailing trend in research studies was to utilize existing conversational assistants such as Amazon Alexa and Google Assistant, with only one paper \cite{4_10.1145/3411764.3445290} opting to create a custom virtual character. And in the post-chatGPT era, there is a notable prevalence of employing chatGPT version 3.5, with a singular paper \cite{11_Bushuven2023} integrating GPT-4.

    \item \textbf{Participant Recruitment Strategy of the Studies:}

    All the studies included child participants with special needs in their research endeavors. Specifically, \cite{6_10.1145/3411764.3445116} focused on recruiting eight Korean adolescents between 16-19 years, where the study involved their collaboration with both caregivers and four experts specializing in Autism Spectrum Disorder (ASD), encompassing counselors, therapists, and clinicians. In contrast, \cite{1_10.1145/3531073.3531157} enlisted three children exhibiting distinct autistic profiles, aged 8, 11, and 9, respectively, where the research interaction involved the children interacting with the vocal interface, with parental assistance provided. In the case of \cite{2_10.1145/3411763.3451666}, the researchers collaborated with nine children diagnosed with neurodevelopmental disorders, aged 4-7, alongside three therapists. \cite{articleGarg} involved 10 students with disabilities (aged 10-15) and five school teachers (one below 30 years, three between 31-40 years, one above 40 years). \cite{4_10.1145/3411764.3445290} conducted research with a sole participant, a 12-year-old child, in collaboration with the child's mother and three other children. Finally, \cite{5_10.1145/3271484} worked with twenty-nine children diagnosed with Autism Spectrum Conditions (ASC) and twelve typically developing children (TD), aged between 4-14. This study involved interactions with a virtual agent and human practitioners, including researchers, teachers, and teaching assistants.

    Concerning research centered on chatGPT, \cite{8_Imran2023}, \cite{9_Yang2023}, and \cite{11_Bushuven2023} did not involve human participants. Instead, \cite{8_Imran2023} concentrated on assisting children, adolescents, and therapists; \cite{9_Yang2023} directed attention to children with Down syndrome, and \cite{11_Bushuven2023} focused on specific queries for chatGPT pertaining to pediatric and life support cases. \cite{7_10.1145/3585088.3593867} enlisted nine interview participants from the United States and South Korea, engaging stakeholders with diverse interdisciplinary backgrounds, including parents, teachers, and AI researchers specializing in computer science. In \cite{10_Moise2023}, the study involved two users and two independent otolaryngologists, with a focus on providing guidance to parents.

    It is noteworthy that prior to the advent of the chatGPT era, all studies enlisted human participants. However, following the introduction of chatGPT, only two studies from our compilation, denoted as \cite{7_10.1145/3585088.3593867} and \cite{10_Moise2023}, engaged human participants. It is evident that, in the post-chatGPT era, researchers transitioned towards a more theoretical exploration of the capabilities of chatGPT in this domain, emphasizing theoretical investigations over practical testing with real participants.

    \item \textbf{Type of Disability Condition and Skill Development:}

    In this specific domain, our focus is directed towards examining the ways in which AI can contribute to the development of specialized skills in children with special needs. Additionally, we scrutinize the predominant types of disability conditions that researchers have primarily addressed within this context.

    The majority of studies during the pre-chatGPT era directed their attention toward children diagnosed with Autism Spectrum Disorder (ASD) \cite{1_10.1145/3531073.3531157}, \cite{2_10.1145/3411763.3451666}, \cite{5_10.1145/3271484}, \cite{6_10.1145/3411764.3445116}, or various other neurodevelopmental disorders, such as global developmental delay \cite{2_10.1145/3411763.3451666}, learning difficulties \cite{5_10.1145/3271484}, and Attention Deficit Hyperactivity Disorder (ADHD) \cite{6_10.1145/3411764.3445116}. Conversely, \cite{articleGarg} and \cite{4_10.1145/3411764.3445290} emphasized physical disabilities, with \cite{4_10.1145/3411764.3445290} addressing blindness and \cite{articleGarg} encompassing learning, visual, hearing, and physical disabilities. In terms of skill development, \cite{1_10.1145/3531073.3531157} focused on assisting children with ASD in acquiring teeth brushing skills, \cite{5_10.1145/3271484} centered on aiding ASD children in developing social communication skills structured around 12 learning activities. Furthermore, \cite{4_10.1145/3411764.3445290} facilitated social sensemaking for blind children in their surrounding situations, while \cite{6_10.1145/3411764.3445116} assisted children with ASD and ADHD in navigating various aspects of daily life, including the promotion of self-care skills, regulation of negative emotions, and practice of conversational skills.

    In the post-chatGPT era, \cite{8_Imran2023} directed attention toward the mental illness of children, emphasizing support for mental healthcare. Concurrently, \cite{9_Yang2023} concentrated on children with Down syndrome, specifically addressing educational benefits, social interaction, overall well-being, and fostering cognitive development while catering to their unique needs. In contrast, \cite{7_10.1145/3585088.3593867} adopted a broader focus on young learners, engaging in visual storytelling to enhance literacy development and encourage creative expression in children. Finally, \cite{10_Moise2023} concentrated on utilizing chatGPT to offer guidance to parents in managing complications arising from post-tympanostomy tube insertion, while \cite{11_Bushuven2023} conducted tests involving chatGPT with twenty pediatric and two basic life support case vignettes.

    Upon analyzing studies from both the pre and post-chatGPT eras, it becomes apparent that researchers predominantly concentrated on Autism Spectrum Disorder (ASD), Attention Deficit Hyperactivity Disorder (ADHD), or various other neurodevelopmental disorders in the pre-chatGPT era. Additionally, some attention was directed towards addressing physical disabilities. The focus was more into promoting daily life skills, or social communication skills. Contrastingly, in the post-chatGPT era, researchers expanded their focus to encompass a diverse array of conditions and skill development areas. This included endeavors such as enhancing children's mental health, fostering creativity, addressing post-medical cases, and promoting cognitive development through the utilization of chatGPT.

\end{enumerate}

\subsection{Overview of Significant Papers from Domain 1:}
In this subsection, we present a brief overview of four significant papers from domain 1. We chose the first 2 papers based on their work significance from the pre-chatGPT era, and the next 2 papers from the post-chatGPT era, and these papers are marked with an asterisk sign in Table \ref{tab:table-1}. We generated the brief summary given all the feature values mentioned in Table \ref{tab:table-1}. 
\begin{itemize}
    \item \textbf{Overview of  \cite{1_10.1145/3531073.3531157}:}
    This study explores the use of ChatGPT 3.5 in conjunction with Amazon Alexa (third-generation Alexa Echo Dot) to assist children with Autism Spectrum Disorder (ASD) in developing teeth brushing skills. The study, conducted at home, involved three children aged 8, 11, and 9, each with different autistic profiles. The evaluation method employed a Semi-Structured Interview with dentists before the study and parents after the study via Zoom. Additionally, an online questionnaire was filled out by parents over seven consecutive days. The paper did not mention any concerns or challenges related to the use of AI. The metrics used for evaluation included Likert scale responses. The study provides insights into the potential of conversational agents to support children with ASD in essential daily tasks.

    \item \textbf{Overview of  \cite{6_10.1145/3411764.3445116}:}
    This work focuses on investigating the efficacy of Voice-based Conversational Agents (VCAs) in aiding adolescents with Autism Spectrum Disorder (ASD) in managing various aspects of their daily lives. The study, based in Korea, involved 8 Korean adolescents aged 16-19. The research utilized an empirical approach, employing interviews (via phone and face to face) as the evaluation method. Although the specific model of the Conversational Agent is not disclosed, the paper falls under the Industrial Design department, emphasizing the Human-Computer Interaction (HCI) aspect. The study, conducted empirically over a two-week period of VCA usage and three workshops in a home environment, raises four concerns: misinterpretation leading to frustration, potential self-esteem issues for adolescents, privacy concerns, and the challenge of sharing the device with family members. The home environment was chosen for the study, and the caregivers played a supporting role. The participants provided informed consent, and an award was given to both adolescents and caregivers (20\$ to each adolescent, and 30\$ to each caregiver in the form of a gift card). The study targeted the development of skills for navigating daily life, addressing the unique challenges faced by individuals with ASD and Attention Deficit Hyperactivity Disorder (ADHD). The data collection involved transcription of recorded audio from interviews and materials from workshops. The study incorporated a Conversational Agent - NUGU CANDLE by SK Telecom, Seoul, Korea, for assistance. However, details regarding voice customization, metrics used, history logs, the number of requests to the voice assistant, and the number of misinterpreted commands were not specified.

    \item \textbf{Overview of  \cite{8_Imran2023}:}
    This research work falls within the domain of psychology and focuses on investigating the potential opportunities and challenges associated with utilizing ChatGPT in child mental healthcare. The study, conducted in Pakistan, is theoretical and does not involve human participants. As there is no human participant, the study does not require any IRB approval or informed consent. The paper raises eight concerns related to the use of ChatGPT which include the presence of bias in trained data, the inability to provide child/adolescent-specific diagnoses, limitations in addressing socioeconomic, educational, and cultural factors, potential provision of inappropriate advice, insensitivity to non-verbal communication, and the incapacity to offer the same level of emotional support as a trained therapist. The research underscores that ChatGPT cannot fully replace human clinical judgment and highlights confidentiality and privacy issues. No specific details regarding the duration of the study, number of sessions, conversational agent, voice customization, metrics, history logs, or the number of requests and misinterpreted commands are provided. Overall, the paper emphasizes caution and outlines areas that require further exploration when incorporating ChatGPT into child mental healthcare.

    \item \textbf{Overview of  \cite{9_Yang2023}:}
    This literature aims to explore the benefits of leveraging ChatGPT in assisting children with Down Syndrome. The study, conducted in Malaysia, is theoretical and does not involve human participants. The paper discusses two concerns related to the use of ChatGPT which revolve around obtaining appropriate, accurate, and sensitive responses for children with Down Syndrome and finding a balance between using ChatGPT and involving educators, therapists, and caregivers. The research underscores the importance of considering the educational benefits, social interaction, overall well-being, cognitive development, and unique needs of children with Down Syndrome. No specific details regarding the duration of the study, number of sessions, conversational agent, voice customization, metrics, history logs, or the number of requests and misinterpreted commands are provided. Overall, the paper emphasizes the potential advantages of ChatGPT in supporting the developmental needs of children with Down Syndrome, while acknowledging key considerations for effective implementation.
  
\end{itemize}

%% file: table-1.tex
\definecolor{Silver}{rgb}{0.752,0.752,0.752}
\begin{longtblr}[
  caption = {Analysis of domain 1},  label={tab:table-1} 
]
{
  width = \linewidth,
  colspec = {Q[31]Q[244]Q[90]Q[204]Q[88]Q[127]Q[146]},
  cells = {c},
  row{2} = {Silver},
  row{9} = {Silver},
  cell{2}{1} = {c=7}{0.929\linewidth},
  cell{9}{1} = {c=7}{0.929\linewidth},
  hlines,
  vlines,
}
\textbf{SL} & \textbf{Paper Name} & \textbf{Diversity of the participa-nts} & \textbf{Evaluation Method} & \textbf{Model of Conversational Chatbot or ChatGPT} & \textbf{Participant Recruitment Strategy} & \textbf{Type of Disability and Skill Development}\\
Pre-ChatGPT Era &  &  &  &  &  & \\
1 & * Alexism: ALEXA supporting children with autism in their oral care at home \cite{1_10.1145/3531073.3531157} & No & Semi structured interview with dentist and parents, Online questionnaire form fillup for parents & Third generation Amazon Alexa echo dot & 3 children with different autistic profiles (8, 11, and 9 years) & Autism Spectrum Disorder (ASD); Teeth Brushing skill\\
2 & Toward the Introduction of Google Assistant in Therapy for Children with Neuro-developmental Disorders: An Exploratory Study \cite{2_10.1145/3411763.3451666} & No & Questionnaire, forms, one hour long semi structured group interview with the therapists & Google Home featured with Google Assistant & 3 therapists, 9 children with NDD (4-7 years) & Neuro-developmental Disorder (ASD, or Global Developmental Delay); No skill development\\
3 & Impact of artificial intelligence in special need education to promote inclusive pedagogy \cite{articleGarg} & Inclusive in terms of disability (not demography) & Focused interviews, content analysis & Not Applicable & 10 students with disabilities (10-15 years), 5 teachers teaching special need children & Learning, Visual, Hearing, and Physical Disability; No skill development\\
4 & Social Sensemaking with AI: Designing an Open-ended AI Experience with a Blind Child \cite{4_10.1145/3411764.3445290} & No & Observational notes, interview from child and mother, email feedback from involved teachers & AR device using Hololens, named PeopleLens & 1 child (12 years) & Visual Disability; Social sensemaking of the surrounding situation\\
5 & Blending Human and Artificial Intelligence to Support Autistic Children’s Social Communication Skills \cite{5_10.1145/3271484} & Inclusive in terms of learning difficulties or ASC (not demography) & Social Communication Questionnaire (SCQ), British Picture Vocabulary Scales (BPVS), 'social communication, emotional regulation and transactional support (SCERTS) model & A virtual character named ANDY & 29 children with ASC, and 12 children with TD (Typically Developing Children) (4-14 years in both group) & Autism Spectrum Conditions, and learning difficulties; Social communication skill\\
6 & * Exploring the Use of a Voice-based Conversational Agent to Empower Adolescents with Autism Spectrum Disorder \cite{6_10.1145/3411764.3445116} & No & Interview~(via phone, and face-to-face) & NUGU CANDLE & 8 adolescents (16-19 years) & Autism Spectrum Disorder (ASD), Attention Deficit Hyperactivity Disorder (ADHD); Navigating various aspects of daily lives\\
Post-ChatGPT Era &  &  &  &  &  & \\
7 & Design implications of generative AI systems for visual storytelling for young learners \cite{7_10.1145/3585088.3593867} & Yes (Participants from both US and South Korea) & Audio interviews (via zoom) & ChatGPT (GPT-3), Dall.E 2, Midjourney, Stable Diffusion & 9 participants (Multiple stakeholders with interdisciplinary backgrounds) & No disability; Visual storytelling for young learners (literacy development and creative expression for children)\\
8 & * Chat-GPT: Opportunities and Challenges in Child Mental Healthcare \cite{8_Imran2023} & No Human Participant & Not Applicable & Not Applicable (only talked about ChatGPT) & Not Applicable & Mental illness; Supporting in mental healthcare\\
9 & * The Potential of ChatGPT in Assisting Children with Down Syndrome \cite{9_Yang2023} & No & Not Applicable & Not Applicable (only talked about ChatGPT) & Not Applicable & Down Syndrome; Promoting cognitive development and supporting unique needs\\
10 & Can ChatGPT Guide Parents on Tympanostomy Tube Insertion? \cite{10_Moise2023} & No & Comparison of generated responses with the recommendations provided by the latest American Academy of Otolaryngology–Head and Neck Surgery Foundation (AAO-HNSF) & ChatGPT 3.5 & 2 users, 2 independent otolaryngologists & \\
11 & ``ChatGPT, can you help me save my child’s life?'' - Diagnostic Accuracy and Supportive Capabilities to lay rescuers by ChatGPT in prehospital Basic Life Support and Paediatric Advanced Life Support cases – an in-silico analysis \cite{11_Bushuven2023} & No Human Participant & Development and validation of the vignettes by 5 emergency physicians & ChatGPT, GPT-4 & Not Applicable & No disability; management of complications post-tympanostomy tube insertion
\end{longtblr}

%% file: Theme2.tex
\section{Analysis of Domain 2 - AI helping women in pregnancy and postpartum period:}
\label{theme:2}

In this section, we present our findings for Domain 2. Table \ref{tab:table-2} represents the significant feature values. We have selected the same features as domain 1, except the last one to be the \textit{Development of a mobile-based Application}, as we consider that the development of a mobile-based application for women in their pregnancy and postpartum period can be a crucial aspect for assisting them.

\input{table-2}

Now, we discuss these dimensions for all of the 10 papers in this particular domain with the analysis implications between the pre and post-chatGPT era:

\begin{enumerate}
    \item \textbf{Diversity of the Participants:}
    
    In the pre-ChatGPT era, studies \cite{12_Ahtisham2023} and \cite{13_Montenegro2022} did not provide details regarding the diversity of their participants in terms of demographic characteristics. Conversely, \cite{14_Mane2023} specifically concentrated on women from communities of color, encompassing individuals belonging to racial or ethnic minority groups, such as African American or Black, Asian, Native American or Alaska Native, Native Hawaiian or other Pacific Islander, and Latinx. The authors of \cite{15_Montenegro2023} asserted their engagement with a diverse participant group.

    Following the chatGPT era, only one paper \cite{16_tsai2023generating} focused on underserved populations, especially those from lower socioeconomic conditions. The authors designed a persona of a pregnant woman belonging to a particular underserved population. On the other hand, \cite{18_Grnebaum2023} did not mention diversity in their work while \cite{17_Venkatasubramanian2022}, \cite{19_Chervenak2023}, \cite{20_Wan2023}, and \cite{21_Sezgin2023} did not involve any human participation.

    It is evident from both pre and post-chatGPT eras that most of the researchers opted out of working with diverse populations, specifically in the post-chatGPT era, lack of human participation was noteworthy.

    \item \textbf{Evaluation Method used in the Research study:}

    In the pre-chatGPT era, the methodologies employed across all studies primarily featured surveys as their principal evaluation tool. For instance, \cite{12_Ahtisham2023} gathered a modest sample of 7 survey responses through platforms such as WhatsApp and Facebook. \cite{13_Montenegro2022} utilized a mixed-methods approach, obtaining qualitative insights from physicians through survey participation, while quantitative data was acquired through surveys distributed to pregnant women via email. In the case of \cite{14_Mane2023}, a post-demonstration survey was administered to solicit feedback concerning the usability, feasibility, and acceptability of their application. \cite{15_Montenegro2023} employed an online survey distributed via email and WhatsApp. In addition to survey instruments, \cite{12_Ahtisham2023} augmented their evaluation with semi-structured interviews involving 7 mothers and 4 gynecologists, with a mix of three online and the rest in-person interviews conducted in the local language. Furthermore, \cite{13_Montenegro2022} conducted structured individual interviews with physicians, capturing responses through both textual and audio means.

    Whereas, in the post-chatGPT era, the research methodologies varied among studies. \cite{16_tsai2023generating} utilized structured questionnaires via Zoom to collect participant feedback, while \cite{17_Venkatasubramanian2022} opted for a non-human participatory approach, comparing results with deep learning models such as LSTM, RNN, and DNN. Conversely, \cite{18_Grnebaum2023} involved physicians in evaluating the responses generated by chatGPT. Studies \cite{19_Chervenak2023}, \cite{20_Wan2023}, and \cite{21_Sezgin2023}, on the other hand, did not incorporate external human participants. In the case of \cite{19_Chervenak2023}, two physicians assessed chatGPT-generated responses to seventeen questions sourced from the FAQ list of the Center for Disease Control and Prevention (CDC) and Fertility knowledge surveys. \cite{20_Wan2023} employed co-authors as a review team, grading chatGPT responses ‘acceptable’ or ‘not acceptable’ based on correctness and completeness in comparison to American College of Obstetricians and Gynecologists (ACOG) publications, PubMed-indexed evidence, and clinical experience. However, they highlighted a significant concern about chatGPT potentially fabricating references and providing non-existent article links and DOIs. Similarly, two authors in \cite{21_Sezgin2023} compared responses from the large language model (LLM) and Google search results to ACOG FAQ responses, rating response quality using a GRADE (Grading of Recommendations Assessment, Development, and Evaluation)-informed scale.

    It is evident that in the pre-chatGPT era, predominant evaluation methods in research studies involved the utilization of survey questionnaires and interviews, often distributed through social media platforms. However, a notable shift in evaluation methods has occurred in the post-chatGPT era. Researchers are increasingly inclined towards assessing responses based on their knowledge or adherence to established medical guidelines.

    \item \textbf{Model of Conversational Chatbot or ChatGPT used in the study:}

    During the pre-chatGPT period, research within this domain predominantly utilized conversational chatbots to provide assistance to women during pregnancy and the postpartum period. Notably, \cite{12_Ahtisham2023} did not explicitly specify the name of the employed chatbot but indicated its use as a daily logger. Conversely, \cite{13_Montenegro2022}, \cite{14_Mane2023}, and \cite{15_Montenegro2023} designed and implemented their custom chatbots, denoted as Maria, Rosie, and BotMaria, respectively. The development process for the chatbot in \cite{13_Montenegro2022} involved the utilization of the DialogFlow tool.

    In the post-chatGPT era, only \cite{17_Venkatasubramanian2022} deviated from working with chatGPT, opting instead for a variation of Generative Adversarial Network (DCGAN) and Convolutional Neural Network (CNN) in their study. Conversely, the remaining papers, \cite{16_tsai2023generating}, \cite{18_Grnebaum2023}, \cite{19_Chervenak2023}, \cite{20_Wan2023}, and \cite{21_Sezgin2023}, integrated chatGPT into their research methodologies. Specifically, \cite{19_Chervenak2023} and \cite{20_Wan2023} utilized chatGPT 3.5, while \cite{21_Sezgin2023} employed chatGPT powered by GPT-4, LaMDA (using BARD), and the Google search engine. However, the models of chatGPT used in \cite{16_tsai2023generating} and \cite{18_Grnebaum2023} were not explicitly mentioned. Notably, \cite{16_tsai2023generating} employed a conversational agent named NutritionBot, which is a GPT-powered AI chatbot developed using Google DialogFlow.

    It is observed that the majority of papers in the pre-chatGPT era chose to develop their proprietary conversational agents. And in the post-chatGPT era, there is a prevalent inclination among researchers to predominantly utilize chatGPT 3.5, as chatGPT version 4 has been introduced just recently.

    \item \textbf{Participant Recruitment Strategy of the Studies:}

    In the pre-chatGPT era, all studies engaged human participants, with a substantial focus on pregnant women and new mothers within this domain. For instance, \cite{12_Ahtisham2023} enrolled 18 women with at least one child under twelve months old, belonging to the upper-middle socio-economic class, equipped with resources for accessing technological devices and possessing moderate digital literacy. \cite{12_Ahtisham2023} also included medical professionals such as gynecologists, psychologists, midwives, and lady health visitors. In \cite{13_Montenegro2022}, 13 pregnant women with an average age of 23 years and 7 physicians aged 28 years participated. \cite{14_Mane2023} concentrated on a diverse group of 109 pregnant women and new mothers of color, aged 14 or older, currently pregnant or parenting a child under 3 years, and belonging to a racial or ethnic minoritized group (African American or Black, Asian, Native American or Alaska Native, Native Hawaiian or other Pacific Islander, and/or Latinx). \cite{15_Montenegro2023} recruited 25 pregnant women aged 18 to 40 years with telephone and internet access through the B.R. Health Centre Cyrio Nacul in the district of Passo Fundo and through referrals. Additionally, \cite{15_Montenegro2023} recruited 10 health professionals, including nutritionists, obstetricians, and family physicians aged 30 and older, using a snowball sampling approach.

    Contrastingly, in the post-chatGPT era, the majority of studies excluded human participants from their research designs. For instance, \cite{19_Chervenak2023} involved two physicians who scrutinized the responses generated by chatGPT in addressing fertility-related clinical prompts against reputable sources. Similarly, in \cite{20_Wan2023}, a review team composed of the co-authors including two board-certified OBGYN physicians, two OBGYN resident physicians, and two second-year medical students (ranging from 24 to 56 years old) evaluated chatGPT responses. Following a similar approach, two authors of \cite{21_Sezgin2023}, both board-certified physicians, compared responses from the Large Language Model (LLM) and Google search results to those from the American College of Obstetricians and Gynecologists (ACOG) FAQ. In \cite{16_tsai2023generating}, three authors, along with one Human-Computer Interaction (HCI) researcher and two graduate students, crafted the persona of a pregnant woman from a lower socioeconomic condition and collaborated with four medical professionals for chatbot co-design. \cite{17_Venkatasubramanian2022} had no human participants, although the dataset comprised 2126 cases of pregnant women between 29 to 42 weeks gestation. In \cite{18_Grnebaum2023}, four physicians formulated the questionnaire, and two physicians posed the questions to chatGPT.

    A notable distinction emerges when comparing the pre and post-chatGPT eras. Research studies conducted before the advent of chatGPT predominantly involved human participants in their investigations. Conversely, in the post-chatGPT era, a shift is observed where many studies opted not to include human participants. Instead, authors directly interacted with chatGPT and assessed its responses based on pre-established guidelines.

    \item \textbf{Development of a mobile-based Application:}

    As this domain focuses on supporting women during pregnancy and addressing the stress and postpartum depression experienced by new mothers, we deem this attribute to be of significant importance. The development of a mobile-based application is perceived as a valuable initiative to aid women in navigating these specific challenges.

    Concerning this specific attribute, an analysis of the pre-chatGPT era reveals that, with the exception of \cite{14_Mane2023}, none of the studies ventured into the development of a mobile-based application. Notably, their application aimed to address racial disparities in maternal and infant health by furnishing new mothers with accessible and trustworthy health information.

    Similarly, in the post-chatGPT era, a parallel pattern emerges. Only \cite{16_tsai2023generating} has embarked on the creation of an application tailored for pregnant women and new mothers, encompassing both mobile and web based platforms. This application is powered by GPT technology and focuses on delivering pregnancy nutrition recommendations.

    It is obvious that in both pre and post-chatGPT era, development of a mobile-based application which can help women in pregnancy and postpartum was not considered as important by the researchers as it seems to be. Rather they focused on working on web based versions of the chatbots, or assessing the responses generated by chatGPT.

\end{enumerate}

\subsection{Overview of Significant Papers from Domain 2}
In this subsection, we present a brief overview of four significant papers from domain 2. We chose the first 2 papers based on their work significance from the pre-chatGPT era, and the next 2 papers from the post-chatGPT era, and these papers are marked with an asterisk sign in Table \ref{tab:table-2}.

\begin{itemize}
    \item \textbf{Overview of  \cite{14_Mane2023}:}
    This research work focuses on the development of Rosie, a free, user-friendly, question-answering chatbot designed to address racial disparities in maternal and infant health. The study is inclusive, targeting a community of color in Washington, District of Columbia, Maryland, and Virginia. The research group contacted community-oriented organizations and participated in events like farmers' markets, citywide festivals, and those offering support to parents through financial or material assistance. This outreach aimed to obtain ongoing feedback on the development of Rosie. The evaluation method involves a post demonstration survey to gather feedback on usability, feasibility, and acceptability of the app. The researchers utilized Rasa's Natural Language Understanding (NLU) and machine learning models. The research, conducted in the field of Epidemiology and Biostatistics, emphasizes community engagement for iterative feedback. The study is empirical, spanning from June to October 2022, with 20 demonstration sessions. IRB approval is not mentioned, but informed consent was obtained from the participants. The target participants include 109 pregnant women and new mothers of color, aged 14 years and older, currently pregnant or parenting a child under 3 years, and part of a racial or ethnic minority group. Data used includes Firebase Database to store and monitor interactions with Rosie, which is a self-developed conversational agent. The study does not mention specific metrics, history logs, the number of requests, or misinterpreted commands. The research resulted in the development of an app rather than a conversational agent capable of speech.

    \item \textbf{Overview of  \cite{15_Montenegro2023}:}
    This study introduces a conversational agent designed to offer reliable information to pregnant women, with a specific focus on nutritional education. The study employs an online survey questionnaire distributed through email and WhatsApp. This study involves the usage of the Doc2Vec model and K-means, along with a neural PV-DM network, emphasizing the Neural Network Architecture for Distributed Memory version of Paragraph Vector (PV-DM). This research is categorized under Public Health, asserts diversity in participants but lacks concrete proof. It is conducted empirically in Brazil from November 2019 to February 2020, with approval from the Brazilian Ethics Committee. The study engaged 35 participants (25 pregnant women from 18 to 40 years old and 10 health professionals having an age of 30 years or older), including pregnant women and health professionals and for each individual, the duration of the experiment was two weeks with their informed consent. The authors employed a range of metrics such as p-value, t-test, mean, cosine similarity, Likert scale, and standard deviation used for assessment. The conversational agent used in this work, named BotMaria, operates as a chat-based interface, not a vocal conversational agent, and historical logs are maintained. The study reports that 48 open messages were left unanswered by this chatbot out of a total of 185 messages.

    \item \textbf{Overview of  \cite{20_Wan2023}:}
    This research work conducts an empirical evaluation of chatGPT as an information source for commonly asked pregnancy questions, focusing on correctness and completeness compared to authoritative references. The study, carried out in the USA, involves query responses graded as ``acceptable" or ``not acceptable" based on American College of Obstetricians and Gynecologists (ACOG) publications, PubMed-indexed evidence, and clinical experience. Utilizing ChatGPT 3.5, the study finds that the model has reference deficiencies for 58\% of the questions and may ``hallucinate" or make up article references or DOI links, potentially causing confusion and stress. The evaluation method includes assessing correctness and completeness, with concerns raised about the potential flaws when used by individuals without medical knowledge. The study did not involve human participants; instead, co-authors reviewed the answers, and the review team consisted of medical professionals including two board-certified OBGYN physicians, two OBGYN resident physicians, and two second year medical students (three males and three females aged from 24 to 56 years old). Moreover, the authors stated that chatGPT generated only one response per query, and they did not assess its reproducibility.

    \item \textbf{Overview of  \cite{21_Sezgin2023}:}
    This literature assesses the quality of responses generated by Large Language Models (LLMs) to frequently asked postpartum depression (PPD) questions in comparison to Google Search responses. The study utilized 14 PPD-related patient-focused frequently asked questions from the American College of Obstetricians and Gynecologists (ACOG). ChatGPT, specifically GPT-4, demonstrated generally higher clinical accuracy in its responses compared to LaMDA using Bard and Google Search Engine. However, it was noted that almost all responses from Bard and ChatGPT lacked source attribution, whereas Google Search results were rated as lower in quality. The evaluation involved no human participants, and the responses were compared by two board-certified physicians, who are also the authors of the study, using a GRADE-informed scale. The study emphasizes the need for improving the sourcing of information in AI-generated responses. The authors employed several statistical tests such as Cohen \textit{k} coefficient, Shapiro-Wilk test, Levene test, Kruskal-Wallis test, and post hoc Dunn test with Bonferroni correction for analysis. No specific duration or number of sessions were mentioned, and the paper did not mention the development of an app.

\end{itemize}

%% file: table-2.tex
\definecolor{Silver}{rgb}{0.752,0.752,0.752}
\begin{longtblr}[
  caption = {Analysis of domain 2},  label={tab:table-2}
]{
  width = \linewidth,
  colspec = {Q[31]Q[185]Q[113]Q[242]Q[100]Q[183]Q[77]},
  cells = {c},
  row{2} = {Silver},
  row{7} = {Silver},
  cell{2}{1} = {c=7}{0.93\linewidth},
  cell{7}{1} = {c=7}{0.93\linewidth},
  hlines,
  vlines,
}
\textbf{SL} & \textbf{Paper Name} & \textbf{Diversity of the participant-s} & \textbf{Evaluation Method} & \textbf{Model of Conversational Chatbot or ChatGPT} & \textbf{Participant Recruitment Strategy} & \textbf{Develop-ment of a mobile-based Application}\\
Pre-ChatGPT Era &  &  &  &  &  & \\
12 & An AI Chat-Based Solution Aimed to Screen Postpartum Depression \cite{12_Ahtisham2023} & No & Surveys (via whatsapp and facebook), and semi-structured interview with mothers and gynecologists (both online and in-person) & Chatbot in the form of a daily logger & 18 women with at least one child less than twelve months old, medical professionals & No\\
13 & Evaluating the use of chatbot during pregnancy: A usability study \cite{13_Montenegro2022} & No & Survey of physicians and pregnant women, Questionnaire (via email) & Maria & 7 physicians (all 28 years old), 13 pregnant women (average age 23 years) & No\\
14 & * Practical Guidance for the Development of Rosie, a Health Education Question-and-Answer Chatbot for New Mothers \cite{14_Mane2023} & Yes (focuses on women from community of color) & Survey & Rosie & 109 pregnant women (14 years and older, currently pregnant or parenting a child younger than 3 years old, and part of a racial or ethnic minoritized group) & Yes\\
15 & * Development and Validation of Conversational Agent to Pregnancy Safe‐education \cite{15_Montenegro2023} & Yes & Online survey questionnaire (via email and whatsapp) & BotMaria & 25 pregnant women (18-40 years), 10 health professionals (all are 30 years and older) & No\\
Post-chatGPT Era &  &  &  &  &  & \\
16 & Generating Personalized Pregnancy Nutrition Recommendations with GPT-Powered AI Chatbot \cite{16_tsai2023generating} & Yes (focused on underserved population, specially with lower socioeconomic condition) & Participant feedback and structured questionnaire (via zoom) & ChatGPT (model not mentioned) & 4 medical professionals & Yes (both mobile and web based)\\
17 & Ambulatory Monitoring of Maternal and Fetal using Deep Convolution Generative Adversarial Network for Smart Health Care IoT System \cite{17_Venkatasubramanian2022} & No human participant & No human participant (result comparison with LSTM, RNN, and DNN) & Not Applicable & No human participant & No\\
18 & The exciting potential for ChatGPT in obstetrics and gynecology \cite{18_Grnebaum2023} & No & Responses evaluated by physicians & ChatGPT (model not mentioned) & 6 physicians & No\\
19 & The promise and peril of using a large language model to obtain clinical information: ChatGPT performs strongly as a fertility counseling tool with limitations \cite{19_Chervenak2023} & No Human Participant & Responses compared with FAQ list of Center for Disease Control and Prevention (CDC), and Fertility Knowledge Surveys by physicians & ChatGPT 3.5 & No human participant & No\\
20 & * ChatGPT: An Evaluation of AI-Generated Responses to Commonly Asked Pregnancy Questions \cite{20_Wan2023} & No Human Participant & Responses compared with American College of Obstetricians and Gynecologists (ACOG) publications, PubMed-indexed evidence, and clinical experience & ChatGPT 3.5 & No human participant & No\\
21 & * Clinical Accuracy of Large Language Models and Google Search Responses to Postpartum Depression Questions: Cross-Sectional Study \cite{21_Sezgin2023} & No Human Participant & Responses and search results compared to the ACOG FAQ responses and rating the quality of responses using a GRADE (Grading of Recommendations Assessment, Development and Evaluation)-informed scale & ChatGPT (GPT-4), LaMDA (using Bard), Google Search Engine & No human participant & No
\end{longtblr}

%% file: Theme3.tex
\section{Analysis of Domain 3 - Ethics and Privacy for Parents while working with AI}
\label{theme:3}

In this section, we present our analysis for Domain 3. Table \ref{tab:table-3} represents the significant feature values. We have selected the same features as domain 1, except the last one to be \textit{Providing suggestions based on the findings}. We acknowledge that as this domain focuses of ethics and privacy while using AI, the researchers should provide future suggestions considering these aspects.

\input{table-3}

Here, we discuss these dimensions for all of the 6 papers in this particular domain with the analysis implications
between the pre and post-chatGPT era:

\begin{enumerate}
    \item \textbf{Diversity of the Participants:}
    
    In the antecedent period predating the advent of ChatGPT, investigations undertaken by \cite{22_10.1145/3025453.3025735} and \cite{23_10.1145/3491102.3502031} involved the engagement of a diverse cohort of participants. In the study conducted by \cite{22_10.1145/3025453.3025735}, the participants were drawn from the Seattle metropolitan area and exhibited a spectrum of socio-economic classes, including both high and low strata, with three families characterized as single-income households. In the study detailed by \cite{23_10.1145/3491102.3502031}, researchers deliberately selected families based on criteria encompassing family structure, ethnicity, geographical location, socio-economic status, children's ages, and gender. Among the 15 families examined, there were five of Asian American and Pacific Islander descent, five classified as White, three identifying as multi-ethnic, and two with Hispanic or Latin affiliation. These families were geographically dispersed across 10 U.S. states in a balanced distribution. Regarding linguistic diversity, 10 of the families reported conversing in languages other than English within their households, encompassing 10 distinct languages and dialects such as Spanish, Chinese, Hindi, Tagalog, Gujarati, and Malayalam. In the case of \cite{24_McStay2021}, the researchers did not explicitly delineate the diversity attributes of the participants, who emanated from non-governmental organizations, industrial sectors, academic institutions, health-related domains, and policymaking realms. Furthermore, \cite{25_feal2020angel} did not involve human participants in its research procedure.

    Nevertheless, during the post-ChatGPT epoch, the investigative undertaking denoted as research study \cite{27_Meyer2023} was devoid of human participant involvement. Conversely, in the context of \cite{26_Luo2023}, researchers engaged participants from both China and the United States.

    Evident within this domain is the observation that researchers in the pre-ChatGPT era demonstrated a greater propensity to engage with a diverse array of participants in comparison to their counterparts in the post-ChatGPT era.

    \item \textbf{Evaluation Method used in the Research study:}

    During the pre-ChatGPT era, a predominant methodology employed by researchers involved the evaluation of their investigative endeavors through the utilization of interviews. For instance, as illustrated by \cite{22_10.1145/3025453.3025735}, in-person semi-structured interviews were conducted with nine child-parent pairs, facilitated by two researchers. Additionally, \cite{24_McStay2021} undertook 13 in-depth interviews and executed a demographically representative national survey within the United Kingdom, comprising 1000 participants, to ascertain parental attitudes towards emotional artificial intelligence in child-centric technologies. The methodology adopted by \cite{23_10.1145/3491102.3502031} entailed a thematic analysis employing an inductive coding framework derived from transcribed video recording sessions, which took place via a video conferencing application. Lastly, \cite{25_feal2020angel} conducted a comprehensive analysis, encompassing both static and dynamic assessments, of the applications under scrutiny.

    Regarding investigations pertinent to the ChatGPT era, \cite{26_Luo2023} conducted semi-structured interviews averaging 65 minutes in duration through online platforms, employing content analysis as their analytical approach. The study adopted the `3A2S' framework, encompassing Accessibility, Affordability, Accountability, Sustainability, and Social Justice considerations, and leveraged the MAXQDA 2022 software for data analysis. In contrast, the study conducted by \cite{27_Meyer2023} primarily focused on discussion surrounding ChatGPT without employing any specific evaluative methodology.

    Both in the pre and post-chatGPT era, we can observe that researchers predominantly employed semi-structured interviews as their primary evaluation methodology. Additionally, in a few instances, surveys were conducted as an alternative assessment modality.

    \item \textbf{Model of Conversational Chatbot or ChatGPT used in the study:}

    During the pre-chatGPT period, the study detailed in \cite{22_10.1145/3025453.3025735} involved the utilization of Hello Barbie and CogniToys Dino as tools for their research. Conversely, in the study denoted as \cite{23_10.1145/3491102.3502031}, participants were instructed to employ the voice assistant already in possession within their households. In instances where a voice assistant was absent, participants were directed to use Siri or download alternative applications such as the Alexa mobile app, Google Voice Assistant, Alexa Voice Assistant, or Asha. In the context of \cite{24_McStay2021}, researchers undertook an examination of various emotional AI toys and conversational agents, including but not limited to Barbie, Kane, and CogniToys Dino. Lastly, the investigative pursuits delineated in \cite{25_feal2020angel} centered on the analysis of 46 parental control applications, diverging from the emphasis on conversational agents as observed in other studies.

    In the post-chatGPT era, only \cite{26_Luo2023} mentioned that they have worked with the chatGPT version 3.5. And researchers in \cite{27_Meyer2023} opted out to mention the version of chatGPT, rather they discussed the opportunities and challenges of chatGPT and LLMs in academia.

   It is noticeable from the analysis that, in the pre-chatGPt era of this domain, researchers preferred working with already existing conversational agents rather than developing new ones.

    \item \textbf{Participant Recruitment Strategy of the Studies:}

    During the era preceding the advent of ChatGPT, with the exception of \cite{25_feal2020angel}, all research endeavors incorporated the recruitment of human participants. For example, \cite{22_10.1145/3025453.3025735} enlisted nine child-parent pairs through a multifaceted approach encompassing email outreach, Facebook posts, and announcements within local parent groups. Eligible parents were specifically chosen based on having children within the age range of 6 to 10 years. In the study detailed by \cite{23_10.1145/3491102.3502031}, researchers recruited a group comprising 18 children aged 5 to 11 years and 16 parents, constituting 15 families. This recruitment transpired through announcements disseminated across various family forums, social media groups, and Slack channels. Furthermore, \cite{24_McStay2021} recruited a total of 13 interviewees and 1000 survey participants for their investigation.

    In the post-chatGPT era, \cite{27_Meyer2023} excluded human participants from their study. Whereas, \cite{26_Luo2023} recruited 4 Chinese professors and 2 American professors who were selected for their expertise in Early Childhood Education (ECE) and proficiency in AI, particularly with a usage history exceeding two months with ChatGPT.  The recruitment was done by leveraging the researchers' interpersonal network.

    In this domain, we can see that most of the participants were recruited via social media and forums in the pre-chatGPT era.

    \item \textbf{Providing Suggestions based on the Findings:}

    In the pre-chatGPT era, each of the studies presented a set of recommendations derived from their respective research findings. Notably, \cite{22_10.1145/3025453.3025735} proposes a reevaluation of the necessity to record and store children's conversations with toys, advocating for the deletion of recordings from both the application and the server after a predetermined duration. Additionally, the study recommends engaging in transparent communication with parents regarding the privacy and security attributes of such toys, positing that this approach may enhance parental consent for their children's engagement with these devices. Furthermore, other works within this category, namely \cite{23_10.1145/3491102.3502031}, \cite{24_McStay2021}, and \cite{25_feal2020angel}, similarly furnish suggestions aimed at addressing ethical and privacy considerations.

    Nevertheless, within the post-ChatGPT era, both studies direct their attention toward recommendations tailored for the academic domain. Specifically, \cite{26_Luo2023} advocates for the integration of these technologies as supplementary tools to augment and fortify the educational experience, positing that ChatGPT can serve as an intelligent assistant for both caregivers and educators. However, a note of caution is sounded, emphasizing that a balanced approach is essential to prevent undue reliance on AI among young children, emphasizing the imperative development of independent critical thinking skills. Similarly, \cite{27_Meyer2023} underscores the potential impediment to students' skill development, such as writing or coding, in the event of excessive dependence on ChatGPT. Nonetheless, the study contends that the central concern should not solely revolve around whether students utilize ChatGPT, but rather, the emphasis should be on discerning how they employ it in their educational endeavors.

    Both in the pre and post-chatGPT era in this domain, we can see a notable amount of suggestions provided by the authors regarding the ethics and privacy issues of using AI and chatGPT, for chatGPT use in the academia, rather than for fostering children.

\end{enumerate}

\subsection{Overview of Significant Papers from Domain 3}
In this subsection, we present a brief overview of two significant papers from domain 3. We chose the 2 papers based on their work significance (1 from pre-chatGPT and 1 from post-chatGPT era) and these papers are marked with an asterisk sign in Table \ref{tab:table-3}.

\begin{itemize}
    \item \textbf{Overview of  \cite{22_10.1145/3025453.3025735}:}
    This study delves into the parental and child reactions to Internet-connected toys, emphasizing mental models and providing recommendations for toy designers, with a primary focus on security and privacy. Originating from the fields of Information Science and Computer Science, the empirical study involved 18 participants (9 parent-child pairs) recruited from the Seattle Metro Area. Employing semi-structured in-person interviews conducted in participants' homes, the study spanned from 40 to 70 minutes per session. While the specific ChatGPT model version used remains unspecified, the study addressed ethical concerns from a parental standpoint, emphasizing privacy and security issues. Obtaining Institutional Review Board (IRB) approval and informed consent, the research highlights suggestions for improvements in toy design and communication with parents regarding privacy.

    \item \textbf{Overview of  \cite{26_Luo2023}:}
    The paper, situated in the ChatGPT era, investigates the optimal roles, challenges, and future developments of ChatGPT in early childhood education (ECE). Conducted through semi-structured interviews with six expert professors from the United States and China, the study explores ethical concerns, biases in training data, plagiarism issues, privacy, and data security. The interviews, lasting from 45 to 79 minutes, were conducted online, and content analysis with inductive and deductive coding methods (using MAXQDA 2022 software) was employed. The participants included four Chinese professors and two American professors, selected based on expertise in ECE and ChatGPT usage. The study emphasizes the ethical implications of AI-generated content, including biases and privacy concerns, and advocates for the establishment of clear ethical and legal frameworks to guide ChatGPT's rational application in education.
  
\end{itemize}

%% file: table-3.tex
\definecolor{Silver}{rgb}{0.752,0.752,0.752}
\begin{longtblr}[
  caption = {Analysis of Domain 3},   label={tab:table-3}
]{
  width = \linewidth,
  colspec = {Q[31]Q[219]Q[188]Q[106]Q[171]Q[123]Q[92]},
  row{even} = {c},
  row{2} = {Silver},
  row{5} = {c},
  row{7} = {Silver,c},
  row{9} = {c},
  cell{1}{1} = {c},
  cell{1}{2} = {c},
  cell{1}{3} = {c},
  cell{1}{5} = {c},
  cell{1}{6} = {c},
  cell{1}{7} = {c},
  cell{2}{1} = {c=7}{0.93\linewidth},
  cell{3}{2} = {c},
  cell{3}{3} = {c},
  cell{3}{4} = {c},
  cell{3}{5} = {c},
  cell{3}{6} = {c},
  cell{3}{7} = {c},
  cell{7}{1} = {c=7}{0.93\linewidth},
  hlines,
  vlines,
}
\textbf{SL}                                                         & \textbf{Paper Name}                                                                                                                          & \textbf{Diversity of the participant-s}                                                                                   & \textbf{Evaluation Method}                                & \textbf{Model of Conversational Chatbot or ChatGPT}                                                          & \textbf{Participant Recruitment Strategy}                               & \textbf{Providing Suggestions based on the Findings} \\
Pre-ChatGPT Era  &                                                                                                                                              &                                                                                                                           &                                                           &                                                                                                              &                                                                         &                                                      \\
22                                                                  & {* Toys that Listen: A Study of Parents, Children, and Internet-Connected Toys\cite{22_10.1145/3025453.3025735}}                                                        & Yes (in terms of socioeconomic classes, no significant differences in demographics)                                       & Semi-structured interviews with parent-child pairs        & Hello Barbie, CogniToys Dino                                                                                 & 18 (9 parent-child pairs, parents with children between 6-10 years)     & Yes                                                  \\
23                                                                  & {Family as a Third Space for AI Literacies: How do children and parents learn about AI together?\cite{23_10.1145/3491102.3502031}}                                    & Yes (in terms of family structure, ethnicity, geographical location, socio-economic background, children ages and gender) & Thematic analysis (study via online)                      & Family owned voice assistant, or Siri, Alexa mobile app, Google voice Assistant, Alexa Voice Assistant, Asha & 18 children (5 to 11 years old) and 16 parents (altogether 15 families) & Yes                                                  \\
24                                                                  & {Emotional artificial intelligence in children's toys and devices: Ethics, governance and practical remedies\cite{24_McStay2021}}                        & No                                                                                                                        & Interviews, survey                                        & Multiple emo toys and CAs designed for children                                                              & 13 interviewee, 1000 survey participants                                & Yes                                                  \\
25                                                                  & {Angel or devil? a privacy study of mobile parental control apps\cite{25_feal2020angel}}                                                                    & No human participant                                                                                                      & Static and dynamic analysis of the apps                   & NA                                                                                                           & No human participant                                                    & Yes                                                  \\
Post-chatGPT Era &                                                                                                                                              &                                                                                                                           &                                                           &                                                                                                              &                                                                         &                                                      \\
26                                                                  & {* Aladdin's Genie or Pandora's Box for Early Childhood Education? Experts Chat on the Roles, Challenges, and Developments of ChatGPT\cite{26_Luo2023}} & Yes (participants from both USA and China)                                                                                & Semi-structured interviews, content analysis (via online) & ChatGPT 3.5                                                                                                  & 4 professors from China, 2 professors from USA                          & Yes                                                  \\
27                                                                  & {ChatGPT and large language models in academia: opportunities and challenges\cite{27_Meyer2023}}                                                        & No human participant                                                                                                      & Not Applicable (only discussed about ChatGPT)             & Not mentioned                                                                                                & No human participant                                                    & Yes                                                  
\end{longtblr}

%% file: Discussion.tex
\section{Discussion}
\label{discussion}

\begin{figure}[ht]
    \center
\includesvg[width=0.7\textwidth]{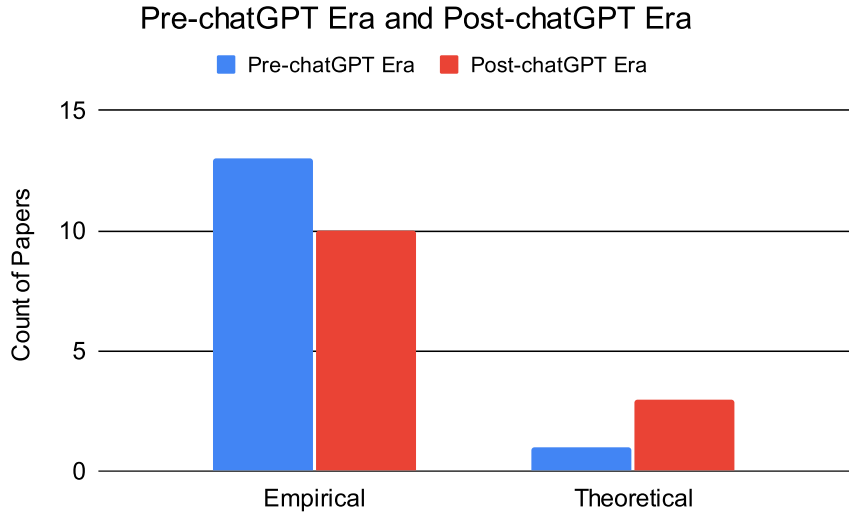}
    \caption{Distribution of Empirical and Theoretical Research Works in the pre and post-chatGPT Era.}
    \label{fig:emp_theo}
\end{figure}

\begin{figure}[ht]
    \center
\includesvg[width=0.7\textwidth]{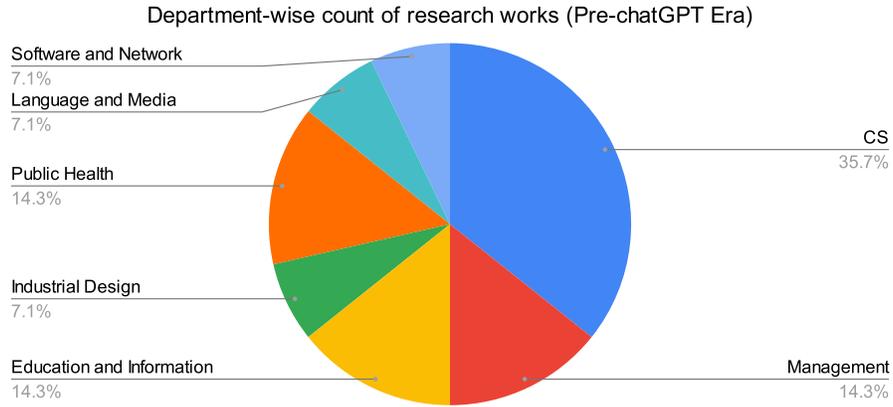}
    \caption{Department-wise distribution of Research Works in the pre-chatGPT Era.}
    \label{fig:dept}
\end{figure}

\begin{figure}[ht]

    \center
\includesvg[width=0.7\textwidth]{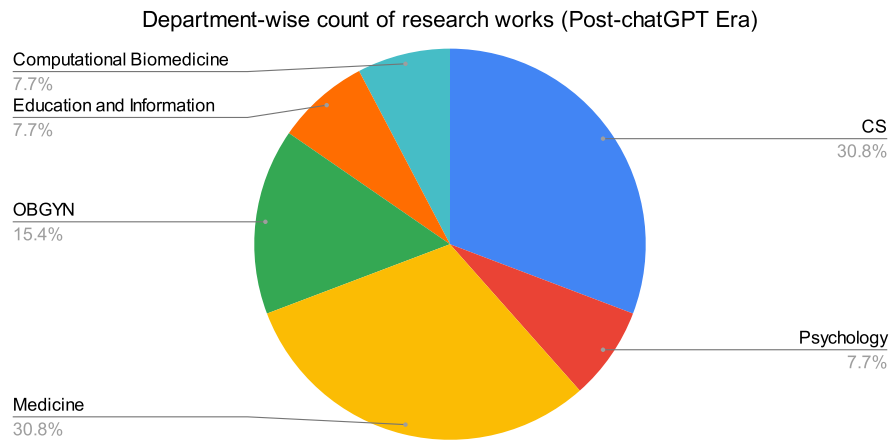}
    \caption{Department-wise distribution of Research Works in the post-chatGPT Era.}
    \label{fig:dept2}
\end{figure}

We present most of our findings from the analysis of the features in Sections \ref{theme:1}, \ref{theme:2} and \ref{theme:3}. However, we try to represent some additional findings in this section.

For example, it is noteworthy and observable that in the pre-chatGPT era, the majority of the research works were empirical. Only 1 research work was theoretical in this timeline. However, in the post-chatGPT era, we can observe a shift in the theoretical work more rather than the empirical ones, which is represented in Figure \ref{fig:emp_theo}.

Also, we can see a shift in the department domain of the research from the pre-chatGPT to the post-chatGPT era. For instance, in the pre-chatGPT era, most of the research works were done by the researchers from the CS community (represented in Figure \ref{fig:dept}). But after the emergence of chatGPT, we can see that, people from other different communities (e.g: Medicine or OBGYN) are also taking a huge part in research incorporating chatGPT, as it is easily accessible by people (represented by Figure \ref{fig:dept2}).

Also, from the analysis of feature 4 for every domain, which is \textit{`Participant Recruitment Strategy'}, it is obvious that most of the studies after the emergence of chatGPT opted out for recruiting human participants, rather the researchers assessed the quality of the chatGPT reponses according to some already published guidelines. Most of the studies in this timeline are conceptual and lacks empirical research.

Moreover, from Figure \ref{fig:work_vs_year}, we can see that a huge upward trend in the research direction for our particular research domain. Specially, after the advent of chatGPT, there is huge scope for research work integrating chatGPT which can help parents to guide and foster their young children in all the three discussed domains.

Lastly, we also find some negative aspects of chatGPT while doing the literature review for our study. For instance, researchers in multiple studies have claimed that chatGPT can generate false or inaccurate information while claiming it with confidence. Also, it can provide the users with fabricated info with fake article references or DOI links. It can create non-existent names and does not provide proper references or citation sources. 

%% file: Conclusion.tex
\section{Conclusion}

In this study, we perform an in-depth analysis of 27 research articles which explores the potential of AI, and LLMs like chatGPT to guide and assist parents in supporting their children. We categorized these 27 papers into 3 domains: \textit{AI helping Special Need Children, AI helping women in pregnancy and postpartum period, Ethics and privacy for parents while working with AI}. Each of the domain was then divided into 2 timeline: pre-chatGPT era, and post-chatGPT era. With a thorough investigation of multiple features for each of the literatures in every criteria, we propose a potential research direction for working with generative AI or large language models along with parents that has great possibilities. Also, we mention the research gaps found upon the analysis, and several negative concerns or aspects of incorporating chatGPT in research dimensions. We also ascertain that after the emergence of chatGPT, most studies are being conceptual and lacks empirical research. However, from our analysis, it can be claimed that chatGPT cannot replace humans or direct interactions between parent-child pair, though it can be integrated to facilitate the bonding and interaction between them.